# GPRS Video Streaming Surveillance System GVSS

**[1]T.P. Pushpavathi R Selvarani[2] N.R. Shashi Kumar[3]**

[1,3]MTech final year, CSE Dept. DSCE
chandravalli@yahoo.co.in, acepushpa@yahoo.co.in
[2]Head Software Engineering Research - RIIC, CSE Department, selvss@yahoo.com
[1,2,3] Dayananda Sagar Institutions, Bangalore, India

**Abstract**

Future security measures will create comfortable living environments that are embedded with a wide range of intelligent functionalities including home computing, entertainment, health care and security. These place stringent requirements on the home networking architecture which integrates various existing technologies for monitoring and control for future high security needs. This paper discusses the design and implementation of a gvss gprs Video Streaming Surveillance System system, which integrates various existing technologies for providing security for smart home environments. This system provides security for office, home and other buildings where high security is required.This allows the mobile user to track the activities from a particular location. The system will send snapshots of the video and stores them in different formats. It is also possible to display the time with the image when it was captured in the gprs enabled mobiles. This system is implemented using J2me Technology.

## I. Introduction

Recent advances in wireless technologies like GSM, Bluetooth, WiFi etc. are coupled with the ability to integrate the multiple requirements of the user to ensure security at their office and home. The possible application areas include smart home, security systems, and patient monitoring [1]. The emergence of several technologies has helped in making intelligent living environments capable of observing and interacting with their surroundings. Such measures assist an user with difficult physical and mental tasks, monitor their safety and health and notify third parties during emergencies. These environments present powerful opportunities for increased independence and a higher quality of living for inhabitants, because personally sensitive information will be used by these systems, but they also pose threats, regarding security problems [2].

The solution to this problem is to create a Security Management System that allows inhabitants to control access to usage, and flow of their personal information. However, there are many challenges in reaching such a goal. From a conceptual perspective, information privacy is a relatively new and complex research area, only being developed in legislative bodies in the last few decades [5]. Further, information privacy is a fluid concept that may change over time as individuals, societies, technologies, and perceptions of privacy change. From a technological perspective, we know that no stable security infrastructure is available to create and test a privacy management system.

Several ubiquitous computing architectures have been proposed, but such systems are yet to be implemented. At present smart home products are proprietary solutions, which are often fairly expensive and provide limited functionality. To overcome these difficulties, we propose to create a model of a GVSS security management system. This will provide a more cost-effective, creative and exhaustive verification of system properties. In this system a mobile phone application using J2ME [4] [5] is designed to control and monitor various types of services. It communicates with a computer server, which is connected to a webcam. In this GVSS system, the owner can view high quality live video recording through a GPRS enabled mobile remotely. The motivation to this GVSS system is, the wireless communication system. in the recent emerging technologies like software system security, Natural language processing, Sensor networks etc. It gives user high quality images with much less GPRS or Universal Mobile Telecommunication System (UMTS) traffic, so communication with the web will not be a cost problem at today's GPRS or UMTS rates[6]. As security is a critical factor for the business and at many places, everyone who utilizes this application can get to know the events happening at a particular location by using a mobile phone from a remote location. This package uses a user friendly Graphical User Interface, people can use it with utmost ease.

## II. Architecture of GVSS

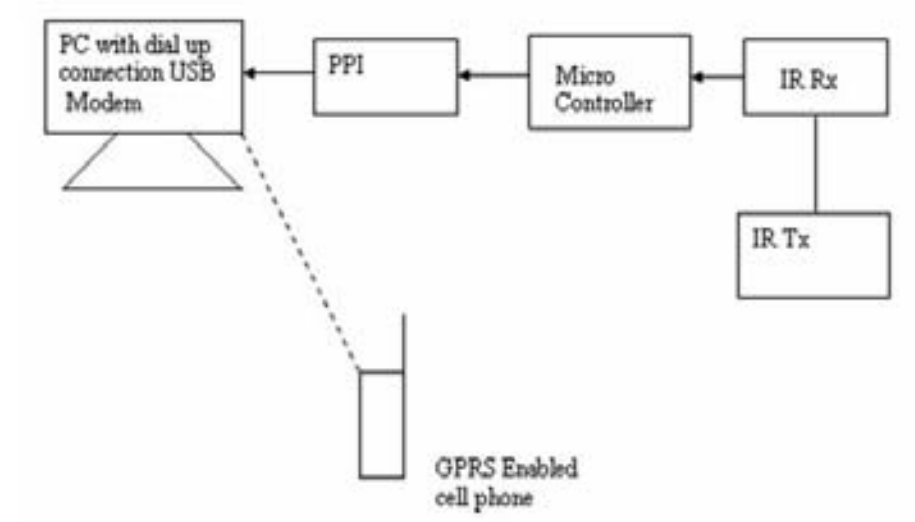

**Figure 1** GVSS architecture

Figure 1 depicts the architecture of the designed system including the various technologies used in this prototype. The system consists of java-enabled mobile phone with a mobile application. A computer server with java application, Infrared transmitter, receiver and web cam are connected to the computer server through a hardware application. In Figure 1 the client network used in the system is a widely used GSM (Global System for Mobile communications) network [3]. A mobile phone communicates with the home server using SMS (Short Message Service) messages. The mobile application, which is written using J2ME (Java 2 Micro Edition) technology includes a menu-driven GUI (Graphic User Interface). Two basic layers of J2ME are used to implement the mobile application, namely CLCD (Connected, Limited Device Configuration) and MIDP (Mobile Information Device Profile). The implemented application using J2ME technology is called a MIDlet. The computer server is the core of the system. The server application is written using java programming language. The system uses Infrared (Ir) Transmitters (Tx) and Receivers (Rx).The Ir transmitter transmits rays continuously to receiver, webcam is placed at the security place and it will be connected to the computer server. The JMF (JavaMedia Framework) is an API (Application Programming Interface), it consists of a suite of three APIs designed for the capture and playback of audio and video. The JMF is helpful for its ease of use, cross platform multimedia, support for different formats, common protocols, extensibility and synchronization of multiple media sources. In the absence of intruders the Ir receiver continuously receives Ir rays, it is monitored by the server. If any intruders are detected when the Ir rays are disturbedand it is detected by the hardware connected to the server. If there is any obstruction, the webcam starts recording the video images and simultaneously an SMS is sent to the authenticated user and the keyboard, mouse of the server is disabled. The system requires the design of two major software components, namely mobile application and the computer server applications. The application of server monitors the Ir rays.

### A. Server model

Figure 2 depicts the server module. It connects the IR hardware at a highly secure place. A Webcam is connected to the server PC through the JMF (java media framework) for capturing images, and it is displayed in a one second delay. Initially "Monitoring Ir"

'.Java program will be running at the server side.

```
Ircheck = new IRCheck ( );
Boolean status = ircheck . checkIRStatus ( );
if(status)
LockingThread lockThread=new LockingThread ( );
SendThread sendThread = new SendThread ( );
CallServer callServer = new CallServer ( );
```

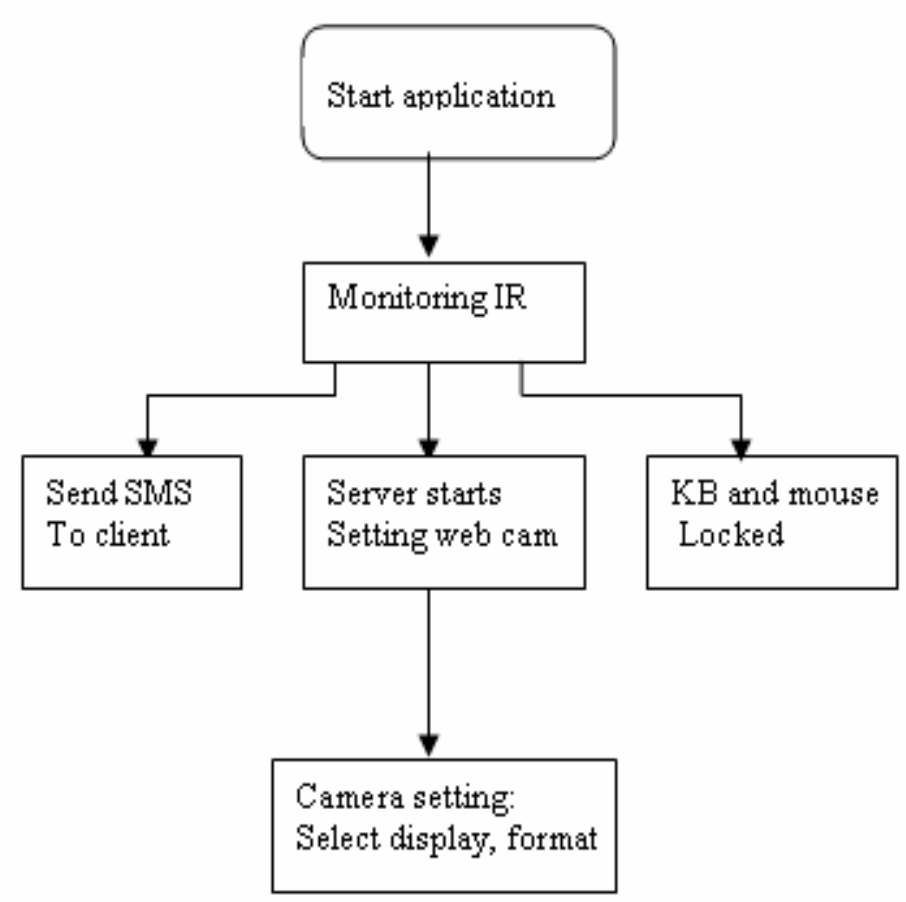

**Figure 2** Server model

It will continuously monitor the IR; if it finds any obstruction it will send an SMS, start the server for transmission of video to the mobile phone and lock the PC. While running the server, the setting window is invoked by using web cam, capturing the videos of specified places, where the web cam is focusing and transferring the video's to the GPRS enabled mobile phone. If there is no interruption between the transmitter and receiver kept at the location where security is needed the receiver will receive rays continuously, and this will be monitored by the application running in the PC–The specified user receives a SMS, and can view the live video feed from their GPRS enabled Mobile, save snapshots, and also unlock the PC (keyboard and mouse).

### B. Mobile model

Figure 3 depicts the model of the mobile application. This module has been implemented by J2ME technology. It is used to take snapshots of video recorded through web cam by connecting to the server. These images are delivered according to each client device capabilities. The server module retrieves the images from the web cam and sends it to the client for display. After launching the application in mobile device, there will be three options:

**View webcam:** Display images from the webcam with the option to save the current image.

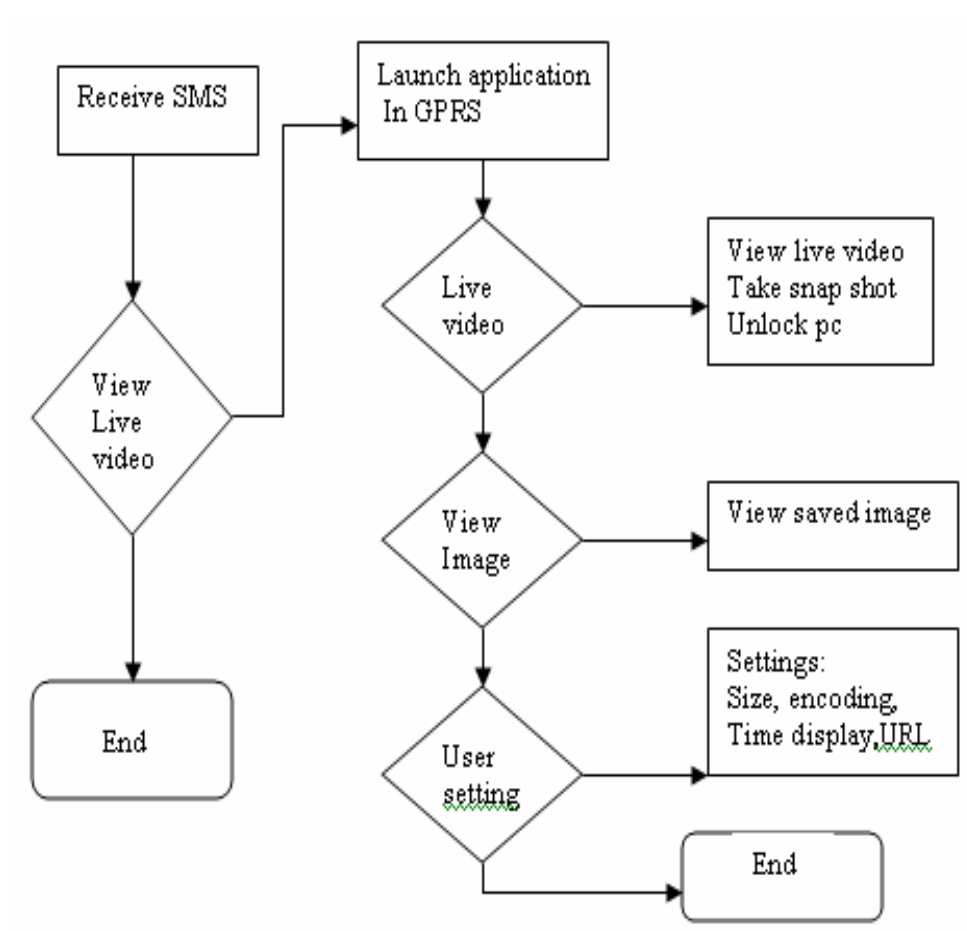

**Figure 3** Mobile model

setSettings( );
midlet.getDisplay(
)setCurrent(midlet.getUserSettingsList( );

**View images:** List of saved images, visualizes or deletes them.
If(imageList == null)
imageList == new ImageList ( this ) ;

**User settings: The** following are available:

**Size:** the image can be scaled to fit into the size provided. If we choose "Constrain Properties" the image will be scaled into the size provided.
ByteArray (getUserSettings( ).
getApproximateImageSize( ) );

**Encoding:** Choose among different image formats such as JPEG, PGN24, PNG 8 or PNG Grayscale.
super.midlet.getUserSettings( ).
setImageEncoding (choice.getSelectedIndex ( ) );

**Time display:** Display the time with the images, and set the display font size.

isDisplayTime( ) and getFontSize ( ) are used.

**URL:** The URL of the webcam server.

connection = (HttpConnection)
Connector.open (loginUrl);

Where Ir Transmitter is connected, and transfers video's to GPRS enabled mobiles. By using the GVSS system security can be monitored. ~~c~~ontinuously by the application running in the PC. The specified user receives SMS's, can view live video from the GPRS enabled Mobile, save snapshots, and also unlock the PC (keyboard and mouse).Figure 4.Describes the model of GVSS System.

GVSS is designed for capturing images and displaying them on J2ME-capable devices. It consists of servlets that runs on a computer with a web cam and a J2ME client. It will display the grabbed video from web cam to the display window. From the video the user can take snapshots and save the images in a mobile device by selecting the save option. The user can unlock the PC by selecting the kill option.

**Kill option:**

```
String Type = request. getParameter ( "Type");
If ((Type).equels ( "Kill" ) ){
Try
{
String exe = "nircmd killprocess " + "WinLock-C.exe"
;
} try
{
Runtime r = Runtime .getRunTime ( );
Process p = r. exec (exe );}}
```

Snapshots can be viewed by selecting the option 'view image'. It will display images from the video and save it. Users can modify the images by encoding, adjusting the image size, as well as display time, by selecting the option user settings. Java Media Framework (JMF) being an Application Programming Interface (API) enables audio, video and other time based media to be added to an application. JMF capture playback enables, stream and transcends multiple media frame formats.

## III. System Services

The GVSS systems main service is to capture images of video recorded through webcam and displaying them on a GPRS enabled mobile. Provides the ability to control any highly secured place remotely. View live video recording through a mobile, and it also allows the mobile user to track the activities happening at a location. The images can be stored in different formats. Also helpful in capturing streaming audio or video in real time using RTP protocol.

Server runs with a fixed public IP or a domain name. It responds to the requests from mobile clients, retrieves video and images from cameras or video clips and passes it to the mobile client. The software is based on Java MIDP 2.0. It works on mobile phones which support Java MIDP 2.0and allow a Java application to access Internet connection and can allocate a Java Application, like Windows Mobile and the PDAs, which have a Java MIDLet Manager, installed. It enables users to select and view cameras in normal or high resolution. It enables users to login with a username and password. It retrieves the list of cameras after login. It allows users to control the camera.

## IV. Conclusions and Future Work

Our novel system GVSS can provide security remotely using a mobile phone. The project provides a wired and wireless system from the end user to the end devices. It integrates a number of technologies to build the complete hardware and software system. The system provides a friendly Graphical User Interface on the mobile phone for ease of use. There are a number of possible extensions to this work such as: support multimedia services through MMS (Multimedia Message Service). Depending upon the coverage area a high-resolution webcam is necessary for clarity of the images. The access management enables parents to have more privileges than children.

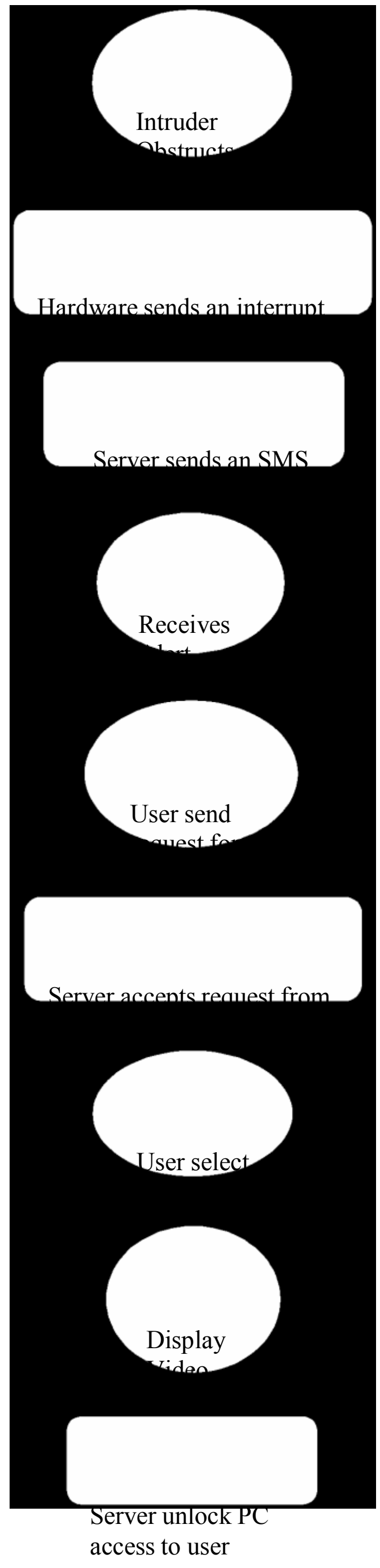

**Figure 4** GVSS Model